\documentclass[aps,prb,reprint,superscriptaddress]{revtex4-1}
\usepackage[utf8]{inputenc}
\usepackage[english]{babel}
\usepackage{amsmath}
\usepackage{float}
\usepackage{amsfonts}
\usepackage{color}
\usepackage{amssymb}
\usepackage{graphicx}

\begin{document}
\title{Role of vertex corrections in the matrix formulation
of the random phase approximation for the multiorbital Hubbard model}

\author{Michaela Altmeyer}
\affiliation{Institut f\"ur Theoretische Physik, Goethe-Universit\"at
Frankfurt,
Max-von-Laue-Stra{\ss}e 1, 60438 Frankfurt am Main, Germany}

\author{Daniel Guterding}
\affiliation{Institut f\"ur Theoretische Physik, Goethe-Universit\"at
Frankfurt,
Max-von-Laue-Stra{\ss}e 1, 60438 Frankfurt am Main, Germany}

\author{ P. J. Hirschfeld}
\affiliation{Department of Physics, University of Florida, Gainesville, FL 32611, USA}

\author{Thomas A. Maier}
\affiliation{Center for Nanophase Materials Sciences and Computer Science and Mathematics Division, Oak Ridge National Laboratory, Oak Ridge, TN 37831-6494, USA}

\author{Roser Valent\'i}
\affiliation{Institut f\"ur Theoretische Physik, Goethe-Universit\"at
Frankfurt,
Max-von-Laue-Stra{\ss}e 1, 60438 Frankfurt am Main, Germany}

\author{Douglas J. Scalapino}
\affiliation{Department of Physics, University of California, Santa Barbara, CA 93106-9530, USA}

\begin{abstract}
In the framework of a multiorbital Hubbard model description of superconductivity,
a matrix formulation of the superconducting pairing interaction that has been widely
used
is designed to treat spin, charge,
and orbital fluctuations within a random phase approximation (RPA). In terms of Feynman
diagrams, this takes into account particle-hole ladder and bubble contributions as
expected.
 It turns out, however, that this matrix formulation also generates additional
terms which have the diagrammatic structure of vertex corrections. Here we examine
these terms and discuss the relationship between the matrix-RPA superconducting
pairing interaction
and the Feynman diagrams that it sums.
  \end{abstract}
  
\maketitle

\section{Introduction}

Despite considerable experimental and theoretical efforts over the last
few decades, unconventional superconductivity remains one of the most interesting
open puzzles in solid-state physics.
While it has been proposed that in some unconventional superconductors
several interactions may be responsible for superconductivity,
 spin fluctuations  are argued by a large fraction of the solid-state
community to be the dominant mechanism driving Cooper pairing
 in heavy-fermion systems, cuprates, two-dimensional organic
charge-transfer salts and the  iron-based
superconductors~\cite{Scalapino2012,Ardavan2012,Hosono2015a,HirschfeldCRAS}.
Such a statement is based on a few
properties that have been found to be characteristic for these material
classes  like the fact that the
superconducting phase is located in close vicinity to an antiferromagnetic
ordered state and the phase transition can be easily tuned by application of
pressure or charge doping~\cite{McKenzie1997, Toyota2007, Tsuei2000,
Stockert2012, Hosono2015a}.

Assuming spin, charge, and orbital fluctuations provide the dominant mechanism driving
superconductivity, various theoretical approaches have been developed in the
last few decades in order to predict superconducting gap functions, critical temperatures, and functional characteristics of thermodynamic quantities.
 Here we shall focus on the random phase approximation (RPA)
for the Hubbard model.
In the single-orbital case on a three-dimensional cubic lattice this approach leads to
a strong enhancement of the
singlet coupling  in the proximity of a spin-density-wave
instability yielding to a  $d_{x^2-y^2}$ symmetry of the gap
function~\cite{Scalapino1986} as observed, for instance,
 in the high-T$_c$
cuprates~\cite{VanHarlingen1995}. Also for superconducting organic charge-transfer
salts~\cite{Powell2006,Saito2012}
this approximation seems to correctly predict the
symmetry of the superconducting gap\cite{Guterding2016,Guterding2016a}.
Recently, such an approximation
has been successfully used
 for the determination of the superconducting gap structure in
iron-based superconductors and heavy-fermion systems~\cite{Takimoto2004,
Kemper2010, Wang2013, Kreisel2013, Suzuki2014, Arai2015, Guterding2015a,
Guterding2015b}.

While the properties of high-T$_c$ cuprates
or organic charge-transfer salts are well captured by single-orbital
models~\cite{Feiner1995, McKenzie1998}, it is crucial to consider multiorbital
models\cite{Takimoto2000,Takimoto2002,Kubo2006,Kuroki2009,Graser2009,Takimoto2004} when aiming for a proper description of heavy-fermion systems
and iron-based superconductors~\cite{Maehira2003, Andersen2011}. Here the Fermi
surfaces consist of several sheets, which emerge from various orbitals at the
Fermi level that participate in the interactions driving  superconductivity.

The matrix-RPA formulation for the single- and multiorbital Hubbard model, like other RPA approaches,
is built around the idea that
 terms involving particle-hole susceptibilities associated
 with a given wave vector $q$ and frequency $\omega $ add up coherently,
while other terms have a {\it random phase} which suppresses their contribution.
Neglection of these {\it random phase} terms in the calculation of the two-particle pairing vertex in the single-orbital case results in a summation of infinite orders of diagrams of pure bubble and ladder topology, yielding an easy-to-evaluate scalar closed-form expression for the pairing interaction.
In the multiorbital case, the susceptibilities acquire a complex orbital dependence, and the interaction strength has to be replaced by full interaction matrices operating on the different orbitals.

 In this work we examine the relation between (i) the diagrammatic representation of
the random phase approximation of the two-particle vertex function in the
context of  the  multiorbital Hubbard model considering
inter- and intraorbital couplings and (ii) the corresponding matrix
representation  of the fluctuation exchange approximation~\cite{Takimoto2004,Kubo2006,Graser2009}.  The latter has been very commonly used to analyze superconductivity in multiband systems, with the assumption that it was equivalent to the usual bubble and ladder diagrams that occur in the one-band RPA. 

In contrast to the derivation based on the separation into the different fluctuation channels~\cite{Zhang2011}, we follow the original diagrammatic approach~\cite{Scalapino1986}
and write down all low-order contributions to the two-particle vertex that can be rewritten as products of the interactions and the bare particle-hole susceptibilities.
Applying this  prescription to the single-orbital case,  we  recover the well-known result for the interaction vertex where only diagrams of pure bubble or ladder topology are incorporated.
In contrast, in the multiorbital case we find that, in addition to the diagrams that have the same topology as in the single-orbital case, there are additional
 diagrams that have the structure of vertex corrections.
We analyze the contributions of such diagrams to certain elements of the pairing interaction and discuss their physical importance for the superconducting state.

\section{The model}
\begin{figure}[tb]
\centering
\includegraphics[width=\columnwidth]{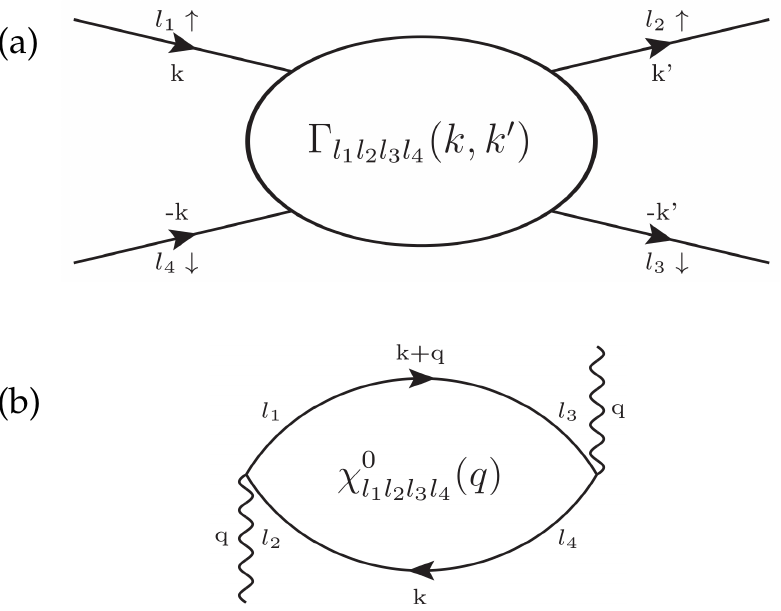}
\caption{Diagrammatic structure of the (a) interaction vertex between two particles with opposite spin and momentum and the (b) matrix elements of the bare susceptibility.}
\label{fig:RPA}
\end{figure}

 Understanding the relationship between the usual diagrammatic
perturbation-resummation approach and the RPA
 matrix formulation provides insight
into the physical processes that are captured in the latter.
 We are interested in relating
the matrix-RPA form of the pairing vertex illustrated in Fig. \ref{fig:RPA}(a) to
its diagrammatic representation.

%For our study it is instructive
 To illustrate our basic point, it is sufficient
to consider a two-orbital problem, where the
noninteracting part of the Hamiltonian $H_0$ is
diagonal in spin $\sigma$ but allows for hopping between the two orbitals
($l=1$ or $2$).   Our discussion can easily be generalized to  more than two orbitals.  The noninteracting Hamiltonian is given by
  \begin{align}
H_0=\sum_{\sigma ll'k} \left(\xi_{ll'}(k) + \epsilon_l \delta_{ll'}\right) d_{l\sigma}^\dagger(k) d_{l'\sigma}^{\phantom{\dagger}}(k).
\label{eq:noninthamil}
\end{align}
For the interacting part of our model Hamiltonian $H_\text{int}$
  we  first restrict ourselves to the intra- and interorbital
 Coulomb interactions $U$ and $U'$, as the effects we wish to discuss are  included:
\begin{align}
H_\text{int}=&U\sum_{k,k',q,l} c_{l\uparrow}^\dagger(k+q) c_{l\downarrow}^\dagger (k'-q) c_{l\downarrow}(k') c_{l\uparrow}(k) \nonumber \\
& +\frac{U'}{2}\sum_{\substack{k,k',q,\\l,l'\neq l,\sigma,\sigma'}} c_{l\sigma}^\dagger (k+q) c_{l'\sigma'}^\dagger (k'-q) c_{l'\sigma'} (k') c_{l\sigma} (k).
\label{eq:inthamil}
\end{align}
We will discuss later what happens when Hund's rule exchange $J$ and pair hopping $J'$ are included.

For this model, the matrix-RPA pairing interaction in the singlet channel derived in the fluctuation exchange (FLEX) approximation~\cite{Takimoto2004,Kubo2006,Kuroki2009, Graser2009} in the vicinity of the critical temperature, where the anomalous Green's function vanishes,
has the following form:
\begin{align}
\begin{split}
\label{eq:singletvertex}
\Gamma_{l_1l_2l_3l_4}^{\textnormal{RPA}} (k,k')=&\left[\frac{3}{2}U^S\chi_1^{\textnormal{RPA}}(k-k')U^S\right.\\
&-\frac{1}{2}U^C\chi_0^{\textnormal{RPA}}(k-k')U^C\\
&\left.+\frac{1}{2}\left(U^S+U^C\right) \right]_{l_1l_2l_3l_4}.
\end{split}
\end{align}
The spin and charge (also commonly called {\it orbital}) interaction matrices
$U^S$ and $U^C$, respectively,
 are $ 4 \times 4 $ matrices with indices $(l_1 l_2)=(11,22,12,21)$. Their
explicit form is given by
\begin{subequations}
\begin{alignat}{3}
U^S=&\begin{pmatrix}
U & 0 & 0 & 0 \\
0 & U & 0 & 0 \\
0 & 0 & U' & 0 \\
0 & 0 & 0 & U'
\end{pmatrix}
\\
U^C=&\begin{pmatrix}
U & 2U' & 0 & 0 \\
2U' & U & 0 & 0 \\
0 & 0 & -U' & 0 \\
0 & 0 & 0 & -U'
\end{pmatrix}
\end{alignat}
\label{eq:interactionmatrices}%
\end{subequations}
The RPA spin ($\chi_1^{\textnormal{RPA}}$) and charge ($\chi_0^{\textnormal{RPA}}$)
 susceptibilities for total spin zero have the form
\begin{subequations}
\begin{align}
\chi_1^{\textnormal{RPA}}(q)=&\left[1-U^S \chi^0(q)\right]^{-1} \chi^0(q)\\
\chi_0^{\textnormal{RPA}}(q)=&\left[1+U^C \chi^0(q)\right]^{-1} \chi^0(q)
\end{align}
\label{eq:rpasusceptibilities}%
\end{subequations}
where the
 matrix elements of the bare susceptibility $\chi^0(q)$ are
given by~\cite{Takimoto2004} (see Fig.~\ref{fig:RPA}(b)):
\begin{align}\label{eq:baresuscep}%
\chi&_{l_1l_2l_3l_4}^0(q,\omega_m) \nonumber \\
&=-\frac{T}{N}\sum_{k,\omega_n}G^0_{l_3l_1}(k+q,\omega_n+\omega_m) G^0_{l_2l_4}(k,\omega_n)
\end{align}
Here $G_{ll'}^0(k,\omega_n)$ denotes the bare Green's function
\begin{align}
 G^0_{ll'}(k,\omega_n) = \sum_\mu \frac{a_\mu^l (k) a_\mu^{l'*}(k)}{i\omega_n-\xi_\mu(k)}
\label{eq:baregreensfunc}
\end{align}
  for an electron with momentum $k$
 and Matsubara energy \mbox{$\omega_n=(2n+1)\pi/\beta$} propagating between the $l$ and $l'$ orbitals.
The weights $a_\mu^l (k)$ are the $l$ elements of the eigenvectors with eigenvalue $\xi_\mu(k)$, which are determined by diagonalization of the noninteracting tight-binding Hamiltonian (Eq.~\ref{eq:noninthamil}).

\begin{figure}[tb]
\centering
\includegraphics[width=\columnwidth]{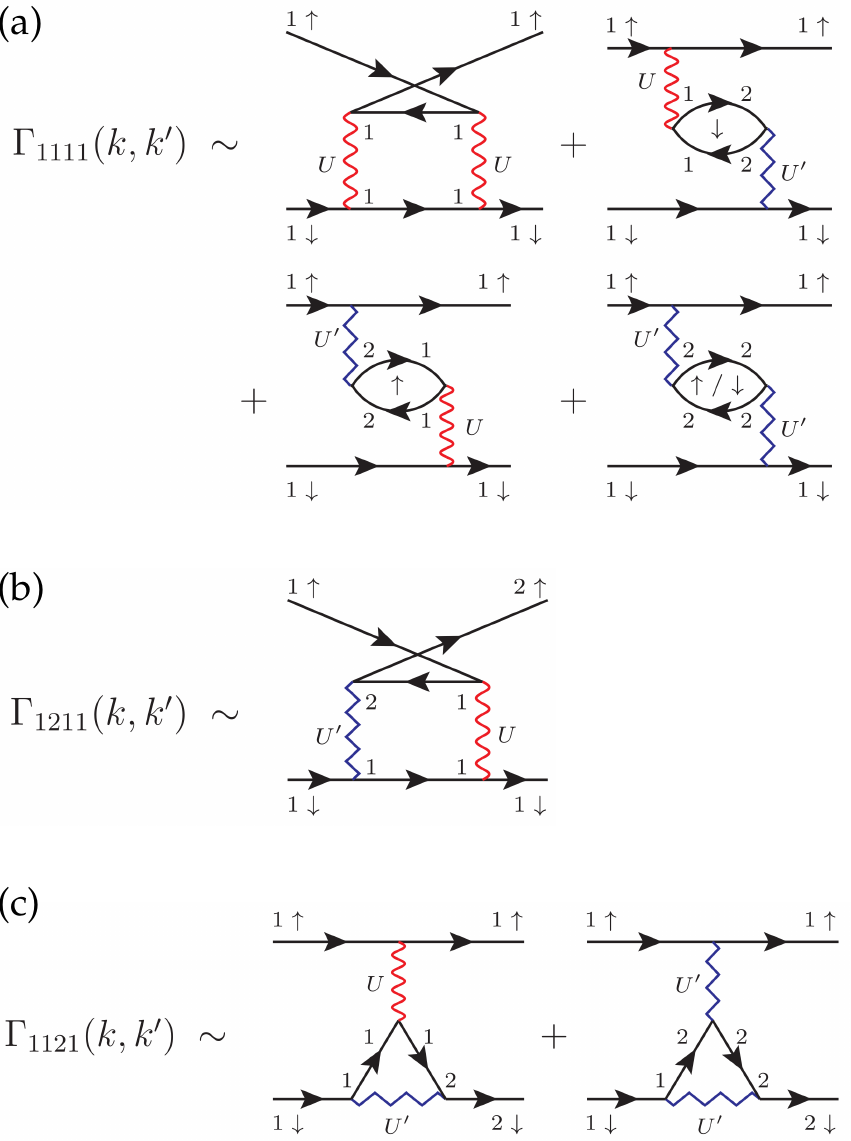}
\caption{Feynman diagrams contributing to (a) $\Gamma_{1111}(k,k')$, (b) $\Gamma_{1211}(k,k')$, and (c) $\Gamma_{1121}(k,k')$. }
\label{fig:combined}
\end{figure}

\section{Results}

We begin  our study of the relationship between the matrix formulation of the RPA
pairing interaction $\Gamma_{l_1l_2l_3l_4}^{\textnormal{RPA}} (k,k')$
 (Eq. \ref{eq:singletvertex}) and the usual diagrammatic
perturbation theory for $\Gamma_{l_1l_2l_3l_4} (k,k')$ by examining two second-order contributions to
$\Gamma_{l_1l_2l_3l_4}^{\textnormal{RPA}}(k,k')$, namely, the
index combinations $\lbrace l_1 l_2 l_3 l_4 \rbrace=\lbrace1111\rbrace$,
\begin{align}\label{eq:Gamma1111}
\Gamma_{1111}^{\textnormal{RPA}}(k,k')\sim \,&  U^2 \chi_{1111}^0(k-k')-UU'\chi_{1122}^0(k-k')\nonumber \\&-UU'\chi_{2211}^0(k-k')-2U'^2 \chi_{2222}^0(k-k'),
\end{align}
and $\lbrace 1121 \rbrace$,
\begin{align}\label{eq:Gamma1121}
\Gamma_{1121}^{\textnormal{RPA}}(k,k')\sim 2UU'\chi_{1121}^0(k-k')+U'^2 \chi_{2221}^0 (k-k').
\end{align}

Using the standard diagrammatic rules for
$\Gamma_{1111}(k,k')$, the last three diagrams shown in Fig.~\ref{fig:combined}(a)
 correspond to the last three terms in Eq. (\ref{eq:Gamma1111})
for $\Gamma_{1111}^{\textnormal{RPA}}(k,k')$.
 The first diagram in Fig.~\ref{fig:combined}(a) gives $U^2\chi_{1111}^0(k+k')$.
 For a singlet and even frequency particle pair
 the Pauli principle implies a certain symmetry of the vertex function:
\begin{align}\label{eq:paulisymmetry}
\Gamma_{l_1l_2l_3l_4}(k,k')=\Gamma_{l_1l_3l_2l_4}(k,-k').
\end{align}

As a consequence, $\Gamma_{1111}(k,k')$ is invariant
 under $k' \to -k'$, and the contribution from the first diagram in
Fig.~\ref{fig:combined}(a) corresponds indeed to the first contribution to the
RPA expression in Eq. (\ref{eq:Gamma1111}).

\begin{figure}[tb]
\centering
\includegraphics[width=0.9\columnwidth]{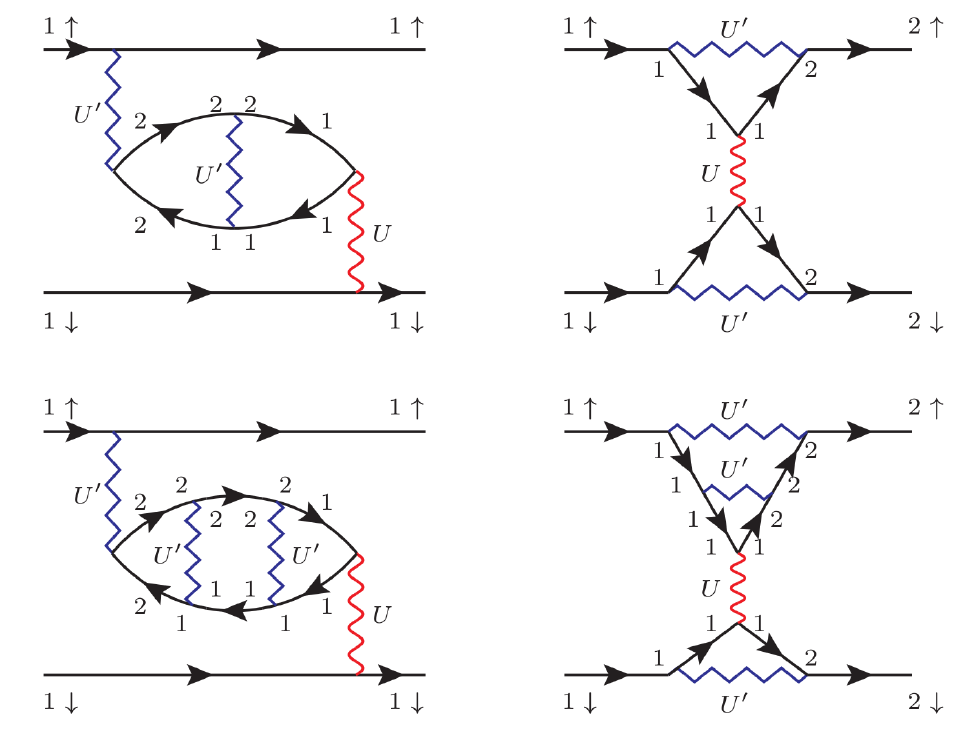}
\caption{Examples of some third- and fourth-order diagrams with $U'$ vertex corrections which contribute to $\Gamma_{1111}$ and $\Gamma_{1221}$. }
\label{fig:vertcorr}
\end{figure}

We investigate now the contributions
 to  $\Gamma_{1121}^{\textnormal{RPA}}(k,k')$ (Eq.~\ref{eq:Gamma1121}) and $\Gamma_{1211}^{\textnormal{RPA}}(k,k')$,
 which are related by symmetry in the singlet channel such that $\Gamma_{1211}(k,k')=\Gamma_{1121}(k,-k')$ (see
Eq.~\ref{eq:paulisymmetry}).  
In the diagrammatic expansion of $\Gamma_{1211}(k,k')$
we have in second order a term of the form
\begin{align}\label{eq:gamma1211}
 UU' \chi_{1121}^0 (k+k'),
\end{align}
as shown in Fig.~\ref{fig:combined}(b).  By symmetry this diagram corresponds to the first term in Eq.
\ref{eq:Gamma1121}:
\begin{align}\label{eq:gamma1211_symmetry}
\Gamma_{1121}^{\textnormal{RPA}}(k,k')\sim UU' \chi_{1121}^0 (k-k').
\end{align}
The remaining contributions to $\Gamma_{1121}^{\textnormal{RPA}}(k,k')$ (Eq. \ref{eq:Gamma1121})
are proportional to $UU'\chi_{1121}^0(k-k')$ and  $U'^2\chi_{2221}^0(k-k')$ respectively.
 As the interorbital Coulomb repulsion preserves
spin and orbital indices at each primitive vertex,
 a scheme based on only bubble and
ladder diagrams leads to no contributions
to $\Gamma_{1121}^{\textnormal{RPA}}(k,k')$.    However,  both terms in Eq. (\ref{eq:Gamma1121})
 can be rewritten
as shown in Fig.~\ref{fig:combined}(c) (left and right diagrams)
 and have the meaning of vertex
corrections in the standard diagrammatic expansion.
 We note that the same type of vertex correction diagrams appear in
functional renormalization-group calculations for the multiorbital Hubbard
model~\cite{Wang2009}.

For higher orders,
 the Feynman diagrams that contribute to
$\Gamma_{l_1l_2l_3l_4}^\textnormal{RPA}$ consist of the familiar bubble and
particle-hole ladder diagrams as well as diagrams that can be
rewritten as vertex corrections arising from $U'$.
Examples of higher-order vertex contributions
are illustrated in Fig. \ref{fig:vertcorr}.

Note that there is a simple way to recognize the Feynman diagrams
that are included in the matrix-RPA:
If one can assign the internal propagators into pairs, such that the  first
propagator is created at the same interaction vertex at which the second
propagator is annihilated and vice versa, then each pair corresponds to the
bare susceptibility as given in Eq. (\ref{eq:baresuscep}) and the contribution
of the considered Feynman diagram to the pairing interaction can be entirely
written in terms of the interactions and the bare susceptibilities.

If a Hund's rule exchange $J$ and a {\it pair hopping} term $J'$ are included
in the multi-orbital interaction (see, e.g.,
Refs.~\onlinecite{Kubo2007,Kemper2010}), the interaction part of the Hamiltonian in real space
is given by
  \begin{align}\label{eq:inthamil_with_hund}
H_\text{int}=&U\sum_{i,l} n_{il\uparrow} n_{il\downarrow} +\frac{U'}{2}\sum_{i,l,l'\neq l} n_{il} n_{il'}\nonumber \\
&+\frac{J}{2}\sum_{\substack{i,l,l'\neq l\\\sigma,\sigma'}} c_{il\sigma}^\dagger c_{il'\sigma'}^\dagger c_{il\sigma'}^{\phantom{\dagger}} c_{il'\sigma}^{\phantom{\dagger}}\nonumber \\
&+\frac{J'}{2}\sum_{i,l, l'\neq l, \sigma} c^\dagger_{il\sigma} c^\dagger_{il\bar{\sigma}} c_{il'\bar{\sigma}}^{\phantom{\dagger}} c_{il'\sigma}^{\phantom{\dagger}},
\end{align}
 and additional vertex terms appear
in the diagrammatic expansion of the pairing interaction.

\begin{figure}[t]
\centering
\includegraphics[width=0.9\columnwidth]{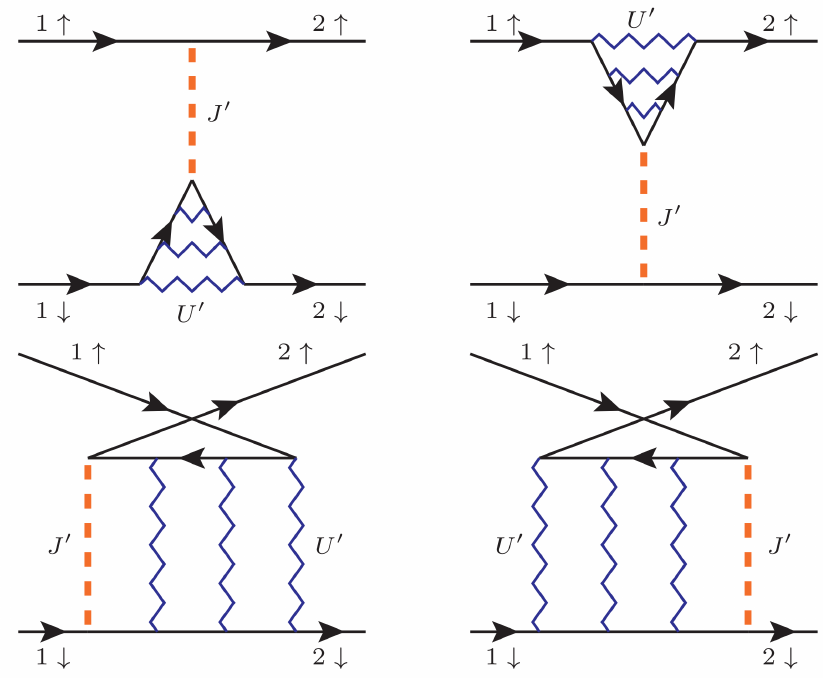}
\caption{When the multiple scattering $U'$ diagrams involving one factor of the pair hopping $J'$ are summed they yield the leading $J'$ result.}
\label{fig:JP}
\end{figure}

We analyze the contributions to the interorbital pair-scattering process
following Ref.~\onlinecite{Kemper2010}: We retain only the diagonal terms of
the bare susceptibilities $\chi_{abab}(q)$ with $a,b=1,2$, assuming that other
terms are suppressed by matrix element effects (see also Eq.
\ref{eq:baregreensfunc}). We find that the dominant second-order contribution
($J\ll U'$) to the pair-scattering term $\Gamma_{abba}$ originates in equal
parts from ladder and vertex diagrams~\cite{comment1}. Summing higher-order terms as shown in
Fig. \ref{fig:JP} we obtain the dominant term  linear in
$J'$:
\begin{align}\label{eq:leadingtermjprime} \Gamma_{1221}^\text{RPA}(k,k')\sim
\frac{2J'U'\chi_{1212}^0(k-k')}{1-U'\chi_{1212}^0(k-k')}+\frac{2J'U'\chi_{2121}^0(k-k')}{1-U'\chi_{2121}^0(k-k')},
\end{align}
which contributes to $\Gamma_{1221}^\textnormal{RPA}(k,k')$ (compare also to the results of the three-orbital matrix-RPA as discussed in Ref.~\onlinecite{Kemper2010}).
For inversion-symmetric systems the two contributions in Eq. (\ref{eq:leadingtermjprime}) are equal, as the bare susceptibility (Eq. \ref{eq:baresuscep}) obeys $\chi_{abcd}^0(q)=\chi_{dcba}^0(-q)$.
A numerical analysis of the importance of these interorbital pair-scattering terms $\Gamma_{abba}$, which
we have shown to contain contributions arising from vertex correction diagrams, for a realistic model of the iron pnictides has recently been conducted by Kemper \textit{et al.}~\cite{Kemper2010}. There, the authors found that in the presence of a nonvanishing pair hopping these terms critically influence the nodal structure of the superconducting gap (see also Fig.~9 of Ref.~\onlinecite{Kemper2010}). 
Moreover, another recent study by Kontani and Onari~\cite{Kontani2010} has shown that exactly the same terms may play a role in enhancing the electron-phonon coupling.

\section{Conclusions}
Based on a simple two-orbital Hubbard model,
we have analyzed the relationship between the
matrix RPA and the Feynman diagram technique for the calculation of
the superconducting pairing interaction. In the single-orbital case, the
 diagrams corresponding to the matrix-RPA
 have been found to have a simple well-known structure containing
only bubble or ladder subunits. 
In the multiorbital case, the interorbital Coulomb interaction  $U'$  and an exchange pair-hopping interaction $J'$ generate additional
diagrams that, as we have shown,
 have the structure of vertex corrections.

The matrix RPA in the context of the multiorbital Hubbard model is therefore a
 significant generalization of the commonly used random phase approximation:
it comprises infinite-order diagrams that
can be written as products of the interactions and the bare susceptibilities.
These include both the usual particle-hole bubble or ladder contributions, and the vertex corrections discussed here, all of which may be summed to infinite order in the usual RPA manner. We have therefore shown that the
commonly used matrix RPA expression  for the pairing vertex in the multiorbital Hubbard model   provides a hitherto unappreciated method to
keep additional contributions to infinite order beyond RPA.
These additional diagrams are important in view of recent studies, as they significantly 
influence the node structure of the superconducting gap
function~\cite{Kemper2010} and may also  enhance phonon-mediated interactions~\cite{Kontani2010}.

\begin{acknowledgments}
We would like to thank K. Zantout, A. R{\o}mer, Y. Wang, P. Lange, P. Kopietz and C. Gros for useful discussions.
M.A., D.G. and R.V. thank the German Research Foundation (Deutsche
Forschungsgemeinschaft) for support through Grants SFB/TR49 and SPP1458.   P.J.H. acknowledges support  through Department of Energy DE-FG02-05ER46236.  T.A.M.  and D.J.S. acknowledge support through the Center for Nanophase Materials Science at ORNL,
which is sponsored by the Division of Scientific User Facilities, U.S. DOE.  M.A. and
R.V. further acknowledge partial support by the Kavli Institute for Theoretical
Physics at the University of California, Santa Barbara under NSF Grant No. PHY11-25915. 
  \end{acknowledgments}

\end{document}